\documentclass[12pt]{article}

\usepackage{graphicx}

\begin{document}

\title{Quark Condensates in Nuclear Matter in  the Global Color
Symmetry Model of QCD\thanks{Work supported by the National
Natural Science Foundation of China and the Foundation for
University Key Teacher by the Ministry of Education}}

\author{{Yu-xin Liu$^{1,2,3,4,5}$, Dong-feng Gao$^{1}$, and Hua Guo$^{2,4}$} \\
\noalign{\vskip 5mm} {\normalsize {$^{1}$ Department of Physics,
Peking University, Beijing 100871, China} }\\
{\normalsize {$^{2}$ The Key Laboratory of Heavy Ion Physics,
Ministry of Education, China, and }} \\
{\normalsize{Department of Technical Physics,
Peking University, Beijing 100871, China} }\\
{\normalsize {$^{3}$ Institute of Theoretical Physics, Academia
Sinica, Beijing 100080, China} }\\
{\normalsize {$^{4}$ Center of Theoretical Nuclear Physics,
National Laboratory of Heavy }}\\
{\normalsize {Ion Accelerator, Lanzhou 730000, China}} \\
{\normalsize{$^5$ CCAST(World Lab.), P. O. Box 8730, Beijing
100080, China}} }

\maketitle


\begin{abstract}
With the global color symmetry model being extended to finite
chemical potential, we study the density dependence of the local
and nonlocal scalar quark condensates in nuclear matter. The
calculated results indicate that the quark condensates increase
smoothly with the increasing of nuclear matter density before the
critical value (about 12$\rho _0$) is reached. It also manifests
that the chiral symmetry is restored suddenly as the density of
nuclear matter reaches its critical value. Meanwhile, the nonlocal
quark condensate in nuclear matter changes nonmonotonously against
the space-time distance among the quarks.
\end{abstract}


\hspace*{2.9cm} Global color symmetry model

\newpage

\parindent=20pt

\section{Introduction}

It is well known that quark condensates are the essential
characteristics in the phase transition from the quark gluon
plasma to hadrons of the early universe evolution. Meanwhile the
condensates feature the nonperturbative structure of the QCD
vacuum, and hence play essential roles in describing hadron
structure, and further the properties of nuclear matter and finite
nuclei\cite{DL901,HP93,Jin95,JK00,SVZ79,RRY85}.  In the aspect of
describing hadron structure in QCD, not only the local scalar
quark condensate $<:\!\bar{q}q\!:>$, but also the nonlocal
condensates $<:\!\bar{q}(x)q(0)\!:>$ are required. Especially, the
later is more important to figure the nonpoint particle property
of hadrons \cite{JK00,JK95,BR91,MR92,WWW01}. Then a large body of
works on the quark condensates in both free space and hot and
dense nuclear matter have existed (for a recent review, see for
example Ref.\cite{DRS01}). Many of the approaches give a descent
feature of the local scalar condensate in nuclear matter at low
energy, which is believed to be the manner for the chiral symmetry
to be restored. However an ``upturn" emerges at higher density in
the linear Walecka model\cite{Mal97}, Dirac-Brueckner
method\cite{LK94} and Schwinger-Dyson formalism\cite{Mit97}.
Recently, the Dyson-Schwinger equation (DSE) method of
QCD\cite{IZ85,RW94} and the global color symmetry model (GCM) of
QCD\cite{CR858,FT9127,Tan97,LLZZ98} have also been employed to
evaluate the condensates in vacuum\cite{Meis97,KM98,Zong99} and
those at finite temperature\cite{JK00} and finite
density\cite{MRS98,BPRS98}. The DSE method
calculations\cite{MRS98,BPRS98} show that the quark condensate
increases with respect to the increase of chemical potential (or
the nuclear matter density) before a critical value is reached. It
means that one has not yet reached a common realization on how the
quark condensate changes against nuclear matter density
qualitatively. Since the DSE approach and the GCM have been shown
to be quite successful in describing hadron properties both in
free space and in nuclear
matter\cite{MRS98,BPRS98,BR96,BRS98,BGP99,HS99,LGG01}, with the
global color symmetry model being extended to that at finite
nucleon density $\rho$ (as usual, it is achieved by involving a
finite chemical potential $\mu$), we will study the dependence of
the local and nonlocal scalar quark condensates on nuclear matter
density again in this paper.

The paper is organized as follows. In Sec. II we describe the
formalism of the GCM  at finite chemical potential $\mu$ and the
relation between the chemical potential and the baryon density in
nuclear matter. In Sec. III we represent the calculation and the
obtained results of the local and non-local scalar quark
condensates as functions of the nuclear matter density. In Section
IV, a brief summary and some remarks are given.

\section{Formalism}

The starting point of the global color symmetry model (GCM) is the
action in Euclidean matric\cite{CR858}
$$ S=\int d^{4}xd^{4}y\left[ \overline{q}(x)\left( \gamma \cdot
\partial + m \right) \delta(x-y) q(y) - \frac{g^2}{2} j^{a}_{\mu}(x)
 D_{\mu \nu}(x-y) j_{\nu}^{a}(y) \right ] \, , $$
 where $j_{\mu}^{a} = \bar{q}(x) \frac{\lambda^a}{2}\gamma_{\mu} q(x)$ is
the quark current, $D_{\mu \nu}(x-y)$ is an effective two-point
gluon propagator, $m$ is the current quark mass, and $g$ is the
quark-gluon coupling constant. Taking the effective gluon
propagator to be diagonal, i.e., $D_{\mu \nu}(x-y) = \delta_{\mu
\nu} D(x-y)$ and applying the Fierz reordering to the quark
fields, one can rewrite the current-current term as
 $$\frac{g^2}{2} \int d^4 x d^4 y j^{a}_{\mu}(x) D(x-y) j_{\nu}^{a}(y) =
- \frac{g^2}{2} \int d^4x d^4y j^{\theta}(x,y) D(x-y)
j^{\theta}(y,x)\, , $$
 where $j^{\theta}(x,y) = \bar{q}(x) \Lambda^{\theta} q(y)$ with
$\Lambda^{\theta}$ being the direct products of Lorentz, flavor
and color matrices which produce the scalar, vector, pseudoscalar
and axial vector terms. More concretely, $\Lambda ^{\theta} =
\frac{1}{2} K^{a} \otimes C^{b} \otimes F^{c}$ with $\{ K^{a} \} =
\{ I, i \gamma_5, \frac{i}{\sqrt{2}} \gamma_{\mu} ,
\frac{i}{\sqrt{2}} \gamma _{\mu} \gamma_{5} \}$, $\{ C^{b} \} = \{
\frac{4}{3} I, \frac{i}{\sqrt{3}} \lambda \}$, $\{ F^{c} \} = \{
\frac{1}{\sqrt{2}} I, \frac{\vec{\tau}}{\sqrt{2}} \}$ where $\{
\lambda ^{a} \} (a=1, 2 , \cdots , 8)$, $\{ \frac{\tau _i} {2} \}
( i = 1, 2, 3) $ are the generators of the groups SU$_{C}$(3) and
SU$_{F}(2)$, respectively. It is obvious that such a
$j^{\theta}(x, y)$ is a bilocal current. With two flavors $u$ and
$d$ of quarks being taken into account, each $\Lambda ^{\theta}$
is either isoscalar or isovector. The color matrices involved in
the Fierz transformation contain color-singlet and color-octet
terms. Taking the bosonization procedure one can transfer the
bilocal quark current structure into auxiliary Bose fields
carrying the quantum number $\theta$. The action of the GCM in
free space (i.e., at the chemical potential $\mu = 0$) for the
zero-mass quark can then be rewritten\cite{CR858} in the Euclidean
space as
$$
S(B)=\int d^{4}xd^{4}y\overline{q}(x)\left[\gamma \cdot \partial
\delta(x-y) + \Lambda^{\theta}B^{\theta}(x,y)\right]q(y) + \int
d^{4}xd^{4}y \frac{B^{\theta}(x,y)B^{\theta}(y,x)}{2g^2D(x-y)} \,
, $$
 where $B^{\theta}(x,y)$ is the bilocal Bose field.

To extend the GCM to finite nuclear matter density (with finite
chemical potential $\mu$), one should, in view of the statistical
mechanics, take the partition function of the canonical (quark)
ensemble into that of the grand ensemble with quarks and hadrons
that are the solitons collecting quarks\cite{FT9127}. The quark
field should be transformed under a constraint on the baryon
number through the chemical potential $\mu$
$$\displaylines{ \hspace*{2cm} q(x) \longrightarrow q^{\prime}(x)
= e^{\mu x_4} q(x) \, . \hfill{(1)} \cr }$$
 After some derivation, we have the action of the GCM in nuclear
matter
$$\displaylines{\hspace*{1cm}  S(B^{\theta}, \mu) = \int \!
d^{4}x d^{4}y \overline{q}^{\prime}(x)\!\left[(\gamma \! \cdot \!
\partial \! - \! \mu\gamma_{4})\delta(x\! - \! y) \! + \!
e^{\mu x_4} \Lambda^{\theta} B^{\theta}(x,y) e^{ -\mu y_4} \right]
q^{\prime} (y) \hfill \cr \hspace*{35mm}  +  \int \! d^{4}xd^{4}y
\frac{B^{\theta}(x,y) B^{\theta}(y,x)} {2g^2D(x-y)} \, , \hfill{}
\cr }$$
 and the generating functional is given as
$$\displaylines{\hspace*{2cm}
Z[\mu , \overline{\eta},{\eta}]=\int D\overline{q} Dq DB^{\theta}
e^{[- S(B^{\theta},\ \mu) + \int \! d^4 x ( \overline{\eta}q +
\overline{q}\eta) ] } \, . \hfill{(2)} \cr }
$$
 where $\overline{\eta}$ and $\eta$ are the quark sources.
After integrating the quark fields, we obtain
$$\displaylines{\hspace*{2mm}
S(B^{\theta}, \mu) \! = \! - \mbox{Tr}{\ln\left[( \gamma\! \cdot
\! \partial \! - \! \mu\gamma_{4}) \delta(x\! - \! y) \! + \!
e^{\mu x_4} \Lambda^{\theta} B^{\theta} e^{-\mu y_4}\right]} \! +
\! \int d^{4}xd^{4}y \frac{B^{\theta}(x,y)B^{\theta}(y,x)}
{2g^2D(x-y)} \, . \hfill{(3)} \cr }$$

Generally, the bilocal field $B^{\theta}(x,y)$ can be written
as\cite{LLZZ98}
$$\displaylines{\hspace*{20mm}
B^{\theta}(x,y) = B^{\theta}_0(x,y) + \sum_{i} \Gamma^{\theta}_{i}
(x,y) \phi^{\theta}_{i}\Big(\frac{x+y}{2} \Big) \, , \hfill{(4)}
\cr }$$
 where $B^{\theta}_{0}(x,y)\! = \! B^{\theta}_0(x-y) = B^{\theta}(x-y)$
is the vacuum configuration of the bilocal field. ${\Sigma_{i}
\Gamma^{\theta}_{i}(x,y) \phi^{\theta}_{i}(\frac{x + y}{2}) }$
correspond to the fields which can be interpreted as effective
meson fields. In the lowest order approximation with only the
Goldstone bosons being taken into account, the $\phi^{\theta}_{i}$
includes $\pi$ and $\sigma$ mesons. The corresponding width of the
fluctuations are approximately the same as the vacuum
configuration\cite{LLZZ98}, i.e., $\Gamma^{\theta}_{i} =
B^{\theta}_{0}$.  Since the bilocal field arises from the bilocal
current of quarks, the internal $\overline{q}$-$q$ structure of
the mesons can be well described in the GCM. The vacuum
configuration can be determined by the saddle-point condition
$\frac{\partial S}{\partial B^{\theta}_0}=0$. Then an equation of
the translation invariant quark self-energy $\Sigma (q,\mu )$ is
obtained as a truncated Dyson-Schwinger equation
$$\displaylines{\hspace*{15mm}
\Sigma(p,\mu)=\int \frac{d^{4}q}{(2\pi)^4}g^2D(p-q)\frac{t^{a}}{2}
\gamma_{\nu} \frac{1}{i\gamma \cdot q-\mu\gamma_{4}+\Sigma(q,\mu)}
\gamma_{\nu} \frac{t^{a}}{2}\, , \hfill{(5)} \cr }$$
 Considering the fact that the inclusion of the chemical potential
breaks the O(4) symmetry in the four-dimensioanl space, one should
rewrite the decomposition of the self-energy $\Sigma$  as
$$\displaylines{\hspace*{20mm}
\Sigma(p,\mu)=i[A(p,\mu)-1]\vec{\gamma} \cdot \vec{p} + i[C(p,\mu)
-1]\gamma_4 (p_4 + i \mu) + B(p,\mu)\, , \hfill{(6)} \cr }
$$
 where $B(p, \mu)$ is the counterpart of the vacuum configuration
in the momentum space, i.e.,
 $$ B(p, \mu) = \int \frac{1}{ (2 \pi)^4} B^{\theta}_{0} (z) e
^{i \tilde{p} z} d z \, .$$
 It requires then that the vacuum configuration of $\sigma$ and $\vec{\pi}$
should satisfy the restriction $\sigma ^2 + \vec{\pi} ^2 = 1 $.
With $\tilde{q}_{\nu}=(\vec{q},q_4+i\mu)$ being introduced, and
combining Eqs.~(5) and (6), one can obtain the equations to
determine the $A(\tilde{p})$, $B(\tilde{p})$ and $C(\tilde{p})$ as
follows
$$\displaylines{\hspace*{15mm}
\left[A(\tilde{p})-1\right]\vec{p}^2 = \frac{8}{3}\int
\frac{d^{4}q}{(2\pi)^4} g^2 D(p-q)\frac{A(\tilde{q})\vec{q}\cdot
\vec{p}} {A^2(\tilde{q}) \vec{q}^2 + C^2(\tilde{q}) \tilde{q}_4^2
+ B^2(\tilde{q})} \, , \hfill{(7a)} \cr \hspace*{15mm}
\left[C(\tilde{p})-1\right]\tilde{p}_4^2 = \frac{8}{3}\int
\frac{d^{4}q}{(2\pi)^4} g^2 D(p-q)\frac{C(\tilde{q})\tilde{q}_4
\tilde{p}_4} {A^2(\tilde{q}) \vec{q}^2 + C^2(\tilde{q})
\tilde{q}_4^2 + B^2(\tilde{q})} \, , \hfill{(7b)} \cr
\hspace*{29mm}
 B(\tilde{p})=\frac{16}{3}\int\frac{d^{4}q}{(2\pi)^4}g^2D(p-q)
 \frac{B(\tilde{q})}{A^2(\tilde{q})\vec{q}^2 + C^2(\tilde{q})
 \tilde{q}^2_4 + B^2(\tilde{q})} \, . \hfill{(7c)} \cr } $$

Basing on the solution of the Dyson-Schwinger equation or the gap
equations, one can determine the bilocal field, and fix further
the GCM action $S$ and the quark propagator $G$ with
$$\displaylines{\hspace*{1cm}
G^{-1} = i A(\tilde{p})\vec{\gamma} \cdot \vec{p} + i C(\tilde{p})
\gamma _4 \tilde{p} _4 + \Lambda^{\theta} B^{\theta} (\tilde{p})
\, . \hfill{(8)} \cr }$$ Thereafter we can evaluate the quark
condensates from the above saddle-point expansion.

It is straightforward that, under the mean-field approximation, for any
quark operator
$$\displaylines{\hspace*{20mm}
{\cal{O}} _n = ( \bar{q}_{j_1} \Gamma^{(1)}_{j_1 i_1} q_{i_1} )
               ( \bar{q}_{j_2} \Gamma^{(2)}_{j_2 i_2} q_{i_2} ) \cdots
               ( \bar{q}_{j_n} \Gamma^{(n)}_{j_n i_n} q_{i_n} )\, ,
\hfill{} \cr }$$
one can get the condensate
$$\displaylines{\hspace*{5mm}
<:{\cal{O}} _n : > = (-1)^n \sum_P \mbox{Tr} [(-)^P
\Gamma^{(1)}_{j_1 i_1}
 \Gamma^{(2)}_{j_2 i_2} \cdots \Gamma^{(n)}_{j_n i_n} (G)_{i_1 jP(1)}
(G)_{i_2 jP(2)} \cdots (G)_{i_n j P(n)} ]\, , \hfill{} \cr }$$
where $G$ is the quark propagator and $P$ denotes the permutation
of the $n$ indices. We obtain then the local scalar quark
condensate as
$$\displaylines{\hspace*{5mm}
<:\bar{q}q:>=<:\overline{q}(0)q(0):>=-\mbox{Tr}_{\gamma C}
{G(x,x)} \hfill{} \cr \hspace*{4.9cm}
 = - 8 \sqrt{2} \int \frac{d^4 p}{(2 \pi)^4}  \frac{B(\tilde{p} ^{2})}
 {\vec{p}^2 A^2(\tilde{p}^2) + \tilde{p}_4^2
C^2(\tilde{p}^2) + B^2(\tilde{p} ^2)} \, , \hfill{(9)} \cr }$$
Meanwhile, the nonlocal quark condensate is given as
$$\displaylines{\hspace*{10mm}
<:\bar{q}(x) q(0) :> = -\mbox{Tr}_{\gamma C} [ G(x,0)]  \hfill{}
\cr \hspace*{36mm} = - 8 \sqrt{2} \int  \frac{d^4 p}{(2 \pi)^4}
\frac{B(\tilde{p}^2)}{\vec{p}^2 A^2(\tilde{p}^2) + \tilde{p}_4^2
C^2(\tilde{p}^2) + B^2(\tilde{p} ^2)} e^{i p x} \hfill{(10)} \cr
}$$

It is apparent that, with the solutions of the gap equations
(7a)-(7c) being taken as the input for Eqs.~(9) and (10), the
variation of the quark condensates against the chemical potential
can be evaluated. In practical calculation, since the knowledge
about the exact behavior of $ g^2$ and $D(p-q)$ in low energy
region is still lacking, one has to take some approximations or
phenomenological form to solve the gap equations. For simplicity,
we adopt the infrared dominant form\cite{MN83,CR858}
$$\displaylines{\hspace*{30mm}
 g^2D(p-q)=\frac{3 \pi ^2 \eta ^2}{q^2} \delta(p-q) \, , \hfill{(11)} \cr } $$
where $\eta$ is an energy-scale parameter and can be fixed by
experimental data of hadrons. Although this form does not include
the contribution from the ultraviolet energy region, it maintains
the main property of QCD in low energy region. With Eqs.~(7a)-(7c)
and (11), one has the Nanbu-Goldstone solution
$$\displaylines{\hspace*{5mm}
A(\tilde{p})=C(\tilde{p})=2, \qquad \qquad
B(\tilde{p})=(\eta^2-4\tilde{p}^2)^{1/2}, \qquad \qquad \mbox{for}
\; \mbox{Re}(\tilde{p}^2)< \frac{\eta^2}{4}, \hfill{(12\mbox{a})}
\cr \hspace*{5mm}
A(\tilde{p})=C(\tilde{p})=\frac{1}{2}\!\left[1\!+\!\left( 1\!+\!
\frac{2\eta^2}{\tilde{p}^2} \right)^{1/2} \right],  \ \
B(\tilde{p})=0, \qquad \qquad \mbox{for} \; \mbox{Re}(\tilde{p}^2)
> \frac{\eta^2}{4}, \hfill{(12\mbox{b})} \cr }  $$
which describes the phase where chiral symmetry is spontaneously
broken and the dressed-quarks are confined. Meanwhile one has also
the Wigner solution
$$\displaylines{\hspace*{10mm}
A(\tilde{p})=C(\tilde{p})=\frac{1}{2}\!\left[1\!+\!\left( 1\!+\!
\frac{2\eta^2}{\tilde{p}^2} \right)^{1/2} \right],  \ \
B(\tilde{p})=0,  \hfill{(13)} \cr }  $$ %
which characterizes a phase in which chiral symmetry is not broken
and the dressed-quarks are not confined.

To fix the energy-scale parameter $\eta$ so that the above
mentioned calculations can be accomplished, we take the way
described in the original work of the GCM\cite{CR858}. Extending
the relation between the GCM and the bag models given in
Ref.\cite{CR858} to that in nuclear matter, we have the relation
among the nucleon radius $R$, nucleon mass $M$ and the bag
constant ${\cal{B}}$ of a nucleon as
$$\displaylines{\hspace*{3cm}
R=\left(\frac{a}{4\pi {\cal{B}}}\right)^{1/4} \, , \hfill{(14)}
\cr }$$
 $$\displaylines{\hspace*{3cm} M = \frac{4a}{3} \left(
\frac{4\pi {\cal{B}}}{a} \right)^{1/4} \, , \hfill{(15)} \cr }$$
where $ a=3(\omega_0 - \mu) - Z_0 =6.12- 3 \mu -Z_0 $, with $Z_0$
being a parameter to include the corrections of the motion of
center-of-mass, zero-point energy, and other effects of the three
quarks in a nucleon. Meanwhile, in advantage of the Friedberg-Lee
nontopological soliton ans\"{a}tz\cite{FL778}, the bag constant
comes from the difference of the action for hadrons and that for
the vacuum. We have then
 $$ {\cal{B}} = S(\mbox{hadron}) - S(\mbox{vacuum}).$$
In the GCM, the interaction is transferred by the Goldstone bosons
$\vec{\,\pi}$ and the scalar meson $\sigma$.  From the restriction
$\vec{\,\pi} ^2 + \sigma ^2 =1$, one can take a simple
approximation $\vec{\,\pi} = 0$, $\sigma =1$ for the vacuum.
Referring the configuration of the mesons in a hadron as $(
\sigma, \vec{\,\pi} )$, we have
$${\cal{B}} = S(\sigma , \vec{\,\pi} ) - S(1,0).$$ %
In principle, the configuration of $\sigma$ and $\vec{\pi}$ should
be determined by solving the coupled equations of motion of the
mesons and the quarks. As an approximation\cite{CR858}, they can
be taken as $\sigma = 0 $, $\vec{\pi} =0$. After some derivation
we obtain
$$\displaylines{\hspace*{5mm}
{\cal{B}}=S(0,0,) - S(1,0) =\! 12 \int \!\!\frac{d^4 p}{(2\pi)^4}
\left\{\ln\left[ \frac{A^{2}(\tilde{p}^2)\vec{p}^2\! + \!
C^{2}(\tilde{p}^2)\tilde{p}_{4}^2\! + \! B^{2}(\tilde{p}^2)}
{A^{2}(\tilde{p}^2)\vec{p}^2 \! + \!
C^{2}(\tilde{p}^2)\tilde{p}_{4}^2 }\right] \right. \hfill{} \cr
\hspace*{6cm}  \left. - \frac{B^{2}(\tilde{p}^2)}
{A^{2}(\tilde{p}^2)\vec{p}^2 \! + \!
C^{2}(\tilde{p}^2)\tilde{p}_{4}^2 \! + \! B^{2}(\tilde{p}^2)}
\right\} \, . \hfill{(16)} \cr }$$

In order to investigate the dependence of the quark condensates on
the nuclear matter density $\rho$ explicitly, we must transfer the
above $\mu$-dependence to the $\rho$-dependence. We should then
determine the relation between the nuclear matter density $\rho$
and the chemical potential $\mu$ in the GCM. It has been known
that the baryon number in nuclear matter can be related with the
generating functional of the system\cite{FT9127,Kap89,Lu02} as
 $$\displaylines{\hspace*{2cm}
 n = \frac{\partial}{\partial \mu} \ln Z(\mu, B^{\theta}(x-y)) \approx
\frac{\partial}{\partial \mu} Tr \ln G^{-1} (\mu , B^{\theta}
(x-y))\, , \hfill{(17)} \cr }$$
 where $G^{-1}(\mu , B^{\theta}(x-y))$ is the inverse of the quark
propagator in the medium. Taking some algorithm, we obtain finally
$$\displaylines{\hspace*{1cm}
n \approx 2 \frac{\partial}{\partial \mu} \int \frac{d^{4}x}{(2
\pi)^{4}} \int d^{4}p \frac{[B^{\theta}(\vec{p}, p_4 + i
\mu)]^2}{p^2} \, . \hfill{} \cr }$$
 The baryon number density can thus be given as
$$\displaylines{\hspace*{1cm}
\rho_{n} = \frac{n}{\int d^{4} x }\approx \frac{2}{(2 \pi)^4}
\frac{\partial}{\partial \mu} \int d^{4}p \frac{[B^{\theta}(
\vec{p}, p_4 + i \mu)]^2}{p^2} \, . \hfill{(18)} \cr }$$

\section{Numerical Result and Discussion}

Taking the energy scale by fitting the properties of $\pi$ and
$\rho$ mesons $\eta = 1.37$~GeV\cite{MRS98} and calibrating the
nucleon mass $M_0=939$~MeV in free space in the GCM, we obtain the
center-of-mass motion correction parameter $Z_0=3.707$ and the
nucleon radius $R_0 = 0.69$~fm. With the fitted parameters and the
above formulae, we get at first, in the Nanbu-Goldstone phase, the
quark condensate in free space $ <:\!\bar{q}q
\!:>_0=-(148~\mbox{MeV})^3$. Meanwhile, in the Wigner phase, the
quark condensate vanishes, i.e., $<:\bar{q}q :>_0 \equiv 0 $. It
is evident that the presently obtained $<:\!\bar{q}q\!:>$ in the
Nanbu-Goldstone phase is very close to the results given with a
fully dressed gluon propagator ($-\!<:\!\bar{q} q\!:>^{1/3}$ can
be $150$-$180$~MeV)\cite{Meis97,MRS98,BRS98} even though it is
smaller than those given in QCD sum rules and some other
approaches. From Eqs.(7) and (10) one can know that the value of
the scalar quark condensate in vacuum depends on the integration
interval determined by the parameter $\eta$. With the increase of
$\eta$, the absolute value of the condensate will increase. In the
present calculation, the $\eta$ is fixed by fitting the nucleon
properties consistently but not freely. What we are now interested
in is the changing characteristic of the condensate in nuclear
matter against the nuclear matter density, which can be identified
as the ratio of the condensates in nuclear matter to that in
vacuum. Since the condensate in nuclear matter depends also on the
parameter $\eta$ with the same relation, the smaller absolute
value in vacuum will not affect the changing feature.

Recalling the property of the quark confinement phase and
deconfinement phase, one can know easily that, in the confinement
phase, the motion of a quark is restricted in a hadron. In a
phenomenological word, the motion of the quark is limited in a
bag. The bag constant provides the constraint to limit the motion
of the quark. It is apparent that, if the bag constant ${\cal{B}}$
becomes zero, the quark can move freely. In view of the variation
of the bag constant, ${\cal{B}}=0$ can then be taken as the
critical point of the confinement phase and the deconfinement
phase. Consequently, the gap equations (7a)-(7c) take the
Nanbu-Goldstone solution or Wigner solution. Accomplishing the
integral in Eq.~(16) with various values of the chemical potential
$\mu$, we get the chemical potential dependence of the bag
constant. With the relation given in Eq.~(18), we obtain the
nuclear matter density dependence of the bag constant. The result
is illustrated in Fig.~1. The figure shows evidently that the bag
constant decreases with respect to the increasing of the nuclear
matter density, and a critical nuclear matter density $\rho\approx
11.8 \rho _{0}$ (corresponding to a critical chemical potential
$\mu _{c} \approx 0.316$~GeV), where $\rho_0$ is the normal
density of nuclear matter, exists for the quark deconfinement to
happen.

By varying the chemical potential $\mu$ and accomplishing the
calculation in Eqs~(9), (10) and (18) at several space-time
distances $x$, we obtain the variation behavior of the ratios of
local and non-local scalar quark condensates in nuclear matter to
the local condensate in vacuum. The results with several fixed
space-time distances are illustrated in Fig.~2 [simply denoted as
$R_2(\rho)$].  The figure shows clearly that, in the
Nanbu-Goldstone phase, the local scalar quark condensate and the
nonlocal condensates with each fixed space-time distance between
the quarks increase smoothly with respect to the increase of
nuclear matter density. As the density of nuclear matter reaches
its critical value $\mu_{c}$, the condensates vanish suddenly. In
the Wigner phase, the condensates maintain zero since
$B(\tilde{p}) \equiv 0$. This result provides a clue that the
chiral symmetry is broken more strongly as the nuclear matter
density increases gradually. When the nuclear matter density
reaches the critical value, the chiral symmetry is restored
suddenly. Such a varying behavior is consistent with the result of
the DSE calculations\cite{MRS98,BPRS98} and the sudden changing
characteristic is similar to that given in the composite-operator
formalism of QCD\cite{Bar90}. However, it is evident that such a
changing behavior of the scalar quark condensates in nuclear
matter is different from most of the previous results. It has been
conventionally known that the nucleon mass is a monotonous
function of the scalar quark condensate, and the mass of a nucleon
in nuclear matter decreases with respect to the increase of
nuclear matter density. Then the absolute value of the scalar
quark condensate should decrease as the nuclear matter density
increases. In fact, the condensate depends also on the number of
nucleons in nuclear matter which increases rapidly against the
increase of nuclear matter density. Combining these two aspects
together, one can know that it is reasonable for the condensates
to increase with respect to the increase of nuclear matter
density. On the other hand, according to the well known
Gell-Mann$-$Oakes$-$Renner relation\cite{GOR68}, the mass of a
pion in nuclear matter will increase with respect to the increase
of nuclear matter density. Consequently, the threshold of the pion
production in ultrarelativistic heavy ion collisions will
increase. One of the reason for the most recently observed
relatively small amount of pion production to proton production in
the RHIC experiment\cite{Seif02} may be such an increase of the
pion production threshold.

Looking over Fig.~2, one may also easily realize that, at a small
space-time distance, the condensate ratios at a definite density
of nuclear matter are larger than that of the local condensate.
However, at large space-time distance, they are smaller. It
indicates that the quark condensate in nuclear matter does not
change monotonously with respect to the space-time distance. To
show this point more explicitly, we display the changing feature
of the ratio of scalar quark condensate in nuclear matter to the
corresponding values in free space against the space-time distance
in Fig.~3 [denoted as $G_2(x^2)$]. The figure shows evidently that
the condensate in free space ($\rho = 0$) decreases gradually with
the increase of the space-time distance among the quarks
definitely. However, the condensate in nuclear matter($\rho \ne
0$) increases if the space-time distance is very small. As the
distance is about $0.2$-$0.24$~fm (about one fourth of the nucleon
radius), the condensate ratio reaches its maximum if the nuclear
matter density changes from $3\rho _0$ to $11\rho _0$, and such a
distance increases with the increase of nuclear matter density. If
the space-timer distance gets larger, the condensate ratio
decreases with the increasing of space-time distance and the
decreasing rate increases with respect to the increase of nuclear
matter density. It manifests that the chiral symmetry can be
restored more rapidly with the increase of nuclear matter density
if the space-time distance among quarks is approximately equal to
or larger than the one forth of the nucleon radius. Furthermore,
such a nonmonotonous changing characteristic also indicates that
there exists a repulsive interaction at very small space-time
distance between the quarks. Since increasing the density of
nuclear matter results generally in a decrease of the distance
among quarks, the decrease of condensate ratios as the quarks are
located farther away beyond the repulsive core is just the other
manifestation of the fact that the absolute values of the scalar
quark condensates increase against increasing nuclear matter
density.

\section{Summary and Remarks}

In summary we have investigated the nuclear matter density
dependence of the local and nonlocal scalar quark condensates in
nuclear matter in an effective field theory model of QCD, namely
the global color symmetry model (GCM). The calculated results
indicate that the quark condensates increase smoothly with the
increase of the density of nuclear matter and vanish suddenly as
the nuclear matter density reaches its critical value (about
$11.8\rho _0$). Meanwhile the nonlocal condensates change
nonmonotonously as the quarks are separated far away from each
other. This result manifests that the chiral symmetry is broken
more seriously as the density of nuclear matter increases and can
be restored suddenly at the critical point. However, if the quarks
are separated farther away from each other, the chiral symmetry
restoration process can be enhanced with the increase of nuclear
matter density.

Since only a very simple form of the $g^2$ and $D(p-q)$ is taken
into account in the present calculation, detailed effects of the
running couple constant $g^2$(or $\alpha$), the gluon propagator
$D(p-q)$ and the other degrees of freedom have not yet been
included. Besides, the dependence of the quark, meson states on
the nuclear matter density (or chemical potential) has not yet
been taken into account self-consistently. Moreover, even though
the presently obtained result is coincident with that obtained in
the DSE formalism\cite{MRS98,BPRS98} and the sudden change feature
is consistent with that given in the composite-operator formalism
of QCD\cite{Bar90}, the changing behavior in the continuous region
is not consistent with most of the previous results (e.g., those
given in Ref.\cite{DRS01}). Then more sophisticated investigation
is necessary and under progress.

\bigskip

This work was supported by the National Natural Science Foundation
of China under Contact No. 10075002, 10135030 and the Major State
Basic Research Development Program under contract No.G2000077400.
One of the authors (YXL) acknowledges the support by the
Foundation for University Key Teacher by the Ministry of
Education, China, too. The authors are also indebted to Professor
Fan Wang, Professor Xiao-fu L\"{u}, Professor En-guang Zhao,
Professor Peng-fei Zhuang, and Professor Hong-shi Zong for their
helpful discussions.

\newpage

\parindent=0pt

\begin{center}
\begin{figure}
\includegraphics[scale=0.8,angle=0]{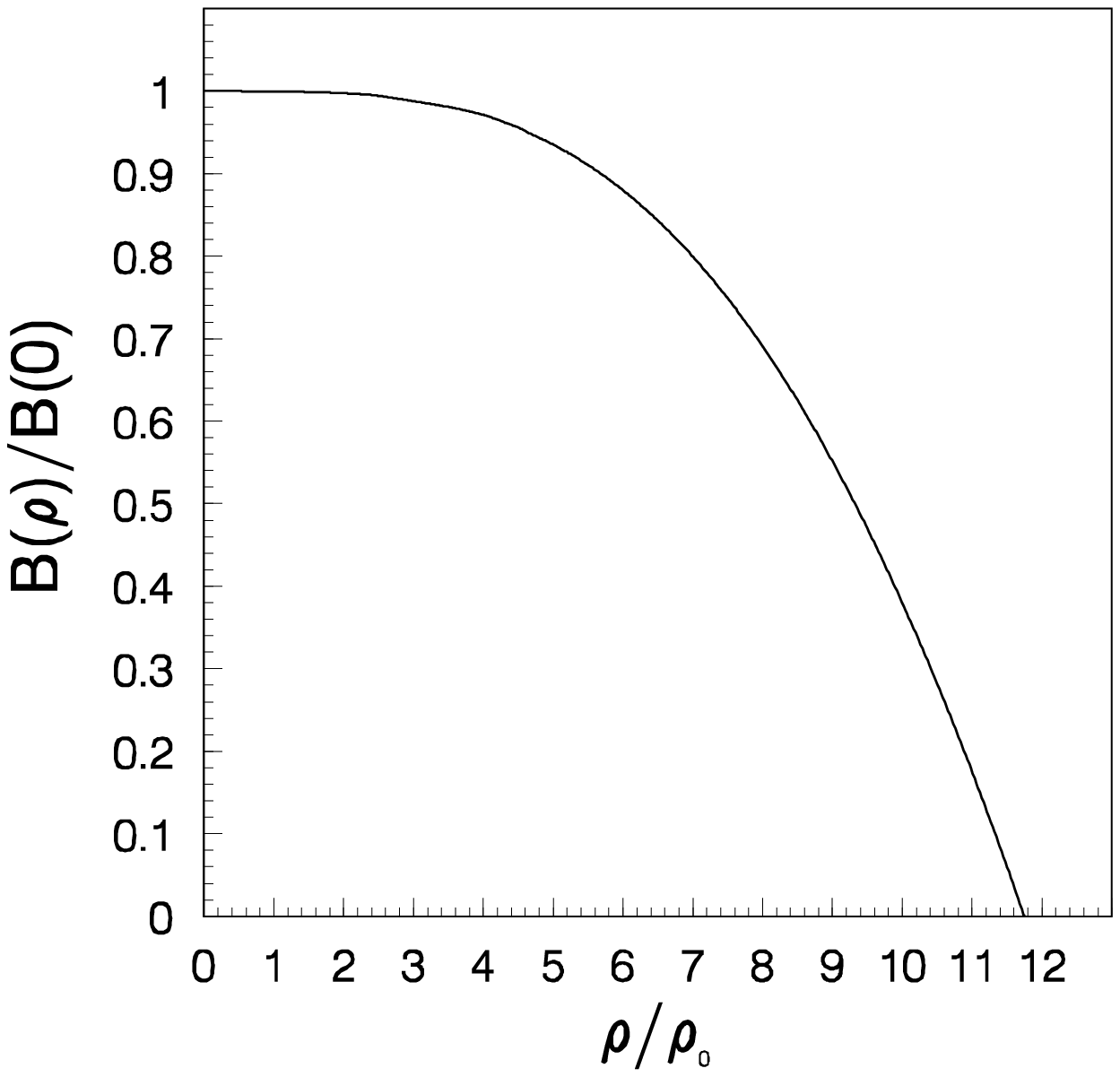}
\caption{The variation of the bag constant with respect to the
nuclear matter density }
\end{figure}
\end{center}

\begin{center}
\begin{figure}
\includegraphics[scale=0.8,angle=0]{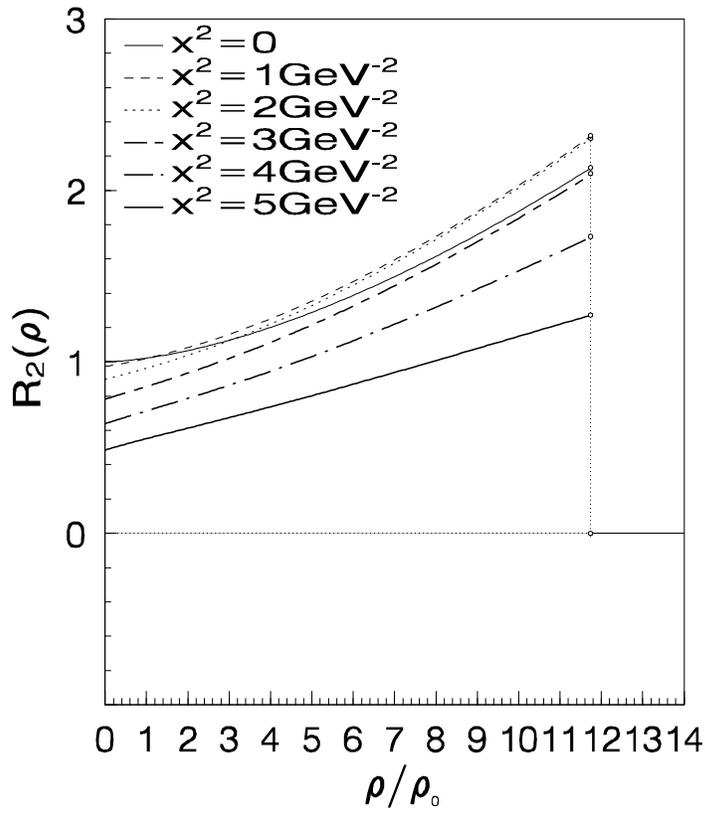}
\caption{Calculated ratio of the two-quark condensate in nuclear
matter to that in free space at several fixed space-time
distances}
\end{figure}
\end{center}

\begin{center}
\begin{figure}
\includegraphics[scale=0.8,angle=0]{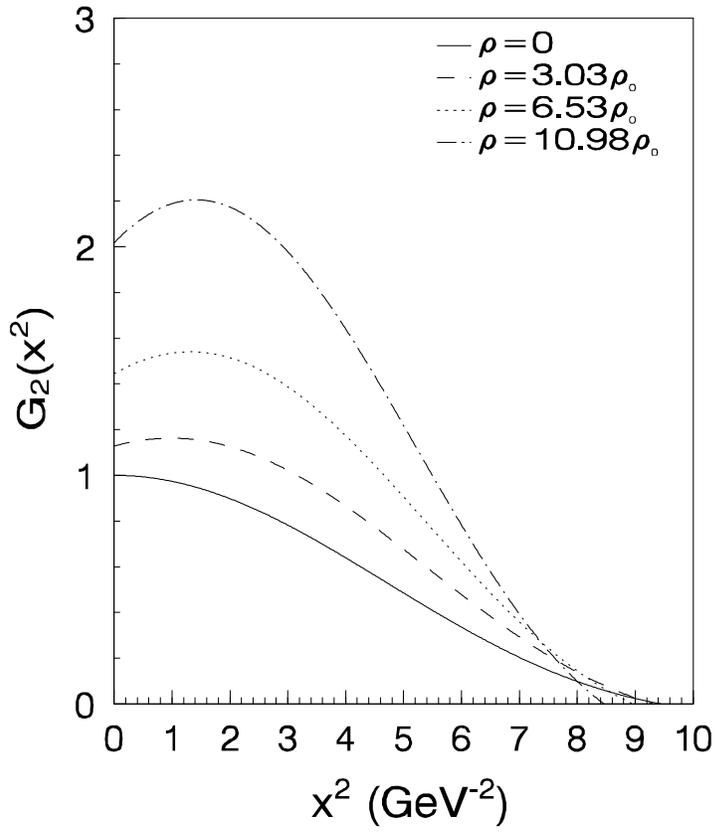}
\caption{Calculated ratio of the nonlocal two-quark condensate in
nuclear matter to that of the local one}
\end{figure}
\end{center}

\end{document}